\documentstyle[12lomcon,cite,graphicx,amssymb]{article}

\newcommand{\bm}[1]{\mbox{\boldmath{$\rm #1$}}}
\newcommand{\ds}{\displaystyle}
\def\roughly#1{\mathrel{\raise.3ex\hbox
{$#1$\kern-.75em\lower1ex\hbox{$\sim$}}}}

\def\gs{\roughly>}
\begin{document}

\vspace*{-3mm}
\begin{flushright}
DESY 06-086\\
June 2006\\
\end{flushright}

\title{NEUTRINOLESS DOUBLE BETA DECAY:\\ ELECTRON ANGULAR CORRELATION \\
AS A PROBE OF NEW PHYSICS}

\author{A.~Ali \footnote{e-mail: ahmed.ali@desy.de}}

\address{Deutsches Electronen-Synchrotron, DESY, 22607 Hamburg, Germany}

\author{ A.V.~Borisov \footnote{e-mail: borisov@phys.msu.ru},
D.V.~Zhuridov \footnote{e-mail: jouridov@mail.ru}}

\address{Faculty of Physics, Moscow State University, 119992 Moscow, Russia}

\maketitle

\vspace*{5mm} \abstracts{ The angular distribution of the final
electrons in the so-called long range mechanism of the
neutrinoless double beta decay ($0\nu2\beta$) is derived for the
general Lorentz invariant effective Lagrangian. Possible theories
beyond the SM are classified from their effects on the angular
distribution, which could be used to discriminate among various
particle physics models inducing $0\nu2\beta$ decays. However,
additional input on the effective couplings will be required to
single out the light Majorana-neutrino mechanism. Alternatively,
measurements of the effective neutrino mass and angular
distribution in $0\nu2\beta$ decays can be used to test the
correlations among the parameters of the underlying
physics models. This is illustrated for the left-right symmetric
model, taking into account current phenomenological bounds.}

\section{Introduction}
Neutrinoless double beta decay ($0\nu2\beta$) is forbidden in the
Standard Model (SM) by lepton number (LN) conservation, which is a
consequence of the renormalizability of the SM. However, being the
low energy limit of a more general theory, an extended version of
the SM could contain nonrenormalizable terms (tiny to be
compatible with experiments), in particular, terms that violate LN
and allow the $0\nu2\beta$ decay. Probable mechanisms of LN
violation can include exchange by: Majorana neutrinos $\nu_M$s
\cite{ZK,Shchep,Doi} (one of the main candidates after the
observation of neutrino oscillations \cite{PDG}), SUSY Majorana
particles
\cite{Mohapatra86,Vergados87,SUSY1,Babu95,SUSY,Faessler97}, scalar
 bilinears (SBs)~\cite{BL}, e.g. doubly charged dileptons (the component
$\xi^{--}$ of the $SU(2)_L$ triplet Higgs scalar etc.),
leptoquarks (LQs) \cite{LQ}, right-handed $W_R$ bosons
\cite{Doi,HKP} etc. From these particles light $\nu$s are much
lighter than the electron and others are much heavier than the
proton, giving rise to the two possible classes of mechanisms for
the $0\nu2\beta$ decay called the long range and the short range
mechanism, respectively.
 For both the classes, the separation of the lepton physics from the nuclear
physics takes place~\cite{Vergados} which simplifies calculations.
 For the first class, in contrast to the second class, the pion exchange mechanism is
suppressed, and in this case operators (in the effective field
theory) stemming from the light neutrino exchange have precisely
the same form as the leading order heavy particle exchange
$0\nu2\beta$ decay operators, enabling a precise comparison among
models~\cite{Prezeau}. According to the Schechter--Valle theorem
\cite{Valle} any mechanism of the $0\nu2\beta$ decay produces an
effective Majorana mass for the neutrino, which must therefore
contribute to this decay in any case. These various contributions
will have to be disentangled to extract information from the
$0\nu2\beta$ decay on the characteristics of the sources of LN
violation, in particular, on the neutrino masses and mixing.

Despite a lack of confirmation for the claimed observation of the
$0\nu2\beta$ decay \cite{Heidelberg}, the restrictions on the
decay half-life \cite{Barabash} make it possible to get bounds on
the parameters of the models with LN violation
(see~\cite{Vergados} for a recent discussion). Once the
$0\nu2\beta$ decay has been established with good accuracy in the
forthcoming experiments, the characteristics of this decay
(half-life and the angular correlation between the electrons)
could be combined with all the updated information from other
experiments on the neutrino mixing and masses (neutrino
oscillations, tritium beta decay \cite{tritium}, cosmology (WMAP
\cite{WMAP03}) etc.) to perform a best fit in the multidimensional
space of parameters of a general underlying particle physics
model. The fit parameters also include the masses and couplings of
the nonstandard particles that could be involved in various
LN-violating processes mentioned  earlier. This would allow to
determine the dominant mechanism (or a set of competing ones) of
the $0\nu2\beta$ decay.

Our aim in this paper is to examine the possibility of the
determination of the decay mechanism from the angular correlation
of the final electrons in this process. We restrict ourselves to
the long-range mechanism and derive the angular distribution for
the general Lorentz invariant effective Lagrangian. The
experimental facilities that can measure the electron angular
distribution in the $0\nu2\beta$ decay are NEMO3 \cite{NEMO3},
ELEGANT V and others \cite{Ejiri}. We argue that the measurement
of the angular correlation coefficient in these experiments would
provide discrimination among the various competing scenarios of
the $0\nu2\beta$ decay. We illustrate this by parametrizing the
angular distribution as $d \Gamma /d \cos \theta \sim 1-K\cos
\theta$ ($K=1$ for the light Majorana mass scenario). Using the
example of the left-right symmetric model \cite{LR}, we work out
the correlation among the angular coefficient $K$, the mass of the
right-handed $W_R$ boson, $m_{W_R}$, and either the effective
Majorana neutrino mass $\langle m\rangle
=\sum\nolimits_{i}U^2_{ei}m_i$ or the half-life $T_{1/2}$, taking
into account the current bounds on the various parameters.
 It is shown that
for values of $\left|\langle m\rangle\right|$ below 10 meV, the
angular correlation between the electrons distinguishes the
left-right symmetric models from the SM $+$ light Majorana mass
scenario.

\section{Angular distribution for the long range mechanism of $0\nu2\beta$ decay}
\subsection{General effective Lagrangian}
For the decay mediated by light $\nu_M$s, the most general
effective Lagrangian is the Lorentz invariant combination of the
leptonic $j_\alpha$ and the hadronic $J_\alpha$ currents of
definite tensor structure and helicity \cite{Limits}
\begin{equation}\label{L}
    {\mathcal L}=\frac{G_F}{\sqrt{2}}[(U_{ei}+\epsilon^{
    V-A}_{V-A,i})j_{V-A}^{\mu
    i}J^+_{V-A,\mu}+\sum\limits_{\alpha,\beta}\!^{^\prime}
\epsilon^\beta_{\alpha i}j^i_\beta J^+_\alpha+{\rm h.c.}]~,
\end{equation}
where the hadronic and leptonic currents are defined as:
$J^+_\alpha=\bar{u} O_\alpha d$ and $j^i_\beta =\bar{e} O_\beta
\nu_i$; the leptonic currents contain neutrino mass eigenstates
and the index $i$ runs over the light eigenstates. Here and
thereafter, a summation over the repeated indices is assumed;
$\alpha$,\,$\beta$=$V\!\mp\!A$,\,$S\!\mp\!P$,\,$T_{L,R}$
($O_{T_\rho}=2\sigma^{\mu\nu}P_\rho$,
$\sigma^{\mu\nu}=\frac{i}{2}\left[\gamma^\mu,\gamma^\nu\right]$,
$P_\rho$ is the projector, $\rho=L,\,R$); the prime indicates the
summation over all Lorentz invariant contributions, except for
$\alpha=\beta=V-A$, and $U_{ei}$ is the PMNS mixing
matrix~\cite{PMNS}. Note that in Eq.~(\ref{L}) the currents
have been scaled relative to the strength of the usual $V-A$
interaction with $G_F$ being the Fermi coupling constant. The
coefficients $\epsilon_{\alpha i}^\beta$ encode new physics,
parametrizing deviations of the Lagrangian from the standard $V-A$
current-current form and mixing of the non-SM neutrinos.

\vspace*{3mm}
In discussing the extension of the SM for the $0\nu2\beta$ decay,
 Ref.~\cite{Doi} considered explicitly only nonstandard terms with
\begin{equation}\label{Doi}
    \epsilon^{V-A}_{V+A,i}=\kappa\frac{g^\prime_V}{g_V}U^\prime_{ei},\quad
\epsilon^{V+A}_{V-A,i}=\eta V^\prime_{ei},\quad
\epsilon^{V+A}_{V+A,i}=\lambda\frac{g^\prime_V}{g_V}V_{ei}~.
\end{equation}
Implicitly, also the contributions encoded by the coefficients
 $\epsilon^{V-A}_{V-A,i}$ are discussed arising from the non-SM contribution
 to $U_{ei}$ in
$SU(2)_L\times SU(2)_R\times U(1)$ models with mirror leptons (see
Ref.~\cite{Doi}, Eq.~(A.2.17)). Here $V$, $U^\prime$ and
$V^\prime$ are the $3\times3$ blocks of mixing matrices for non-SM
neutrinos, e.g., for the usual $SU(2)_L\times SU(2)_R\times U(1)$
model $V$ describes the lepton mixing for neutrinos from
right-handed lepton doublets; for $SU(2)_L\times SU(2)_R\times
U(1)$ model with mirror leptons \cite{Pati} $U^\prime$
($V^\prime$) describes the lepton mixing for mirror
left(right)-handed neutrinos \cite{Doi} etc. The form factors
$g_{V}$ and $g^\prime_V$ are expressed through the mixing angles
for left- and right-handed quarks. Thus,  $g_V=\cos \theta_C$ and
$g_V^\prime = e^{i \delta} \cos \theta_C^\prime$, with $\theta_C$
being the Cabibbo angle, $\theta_C^\prime$ is its right-handed
mixing analogoue, and the CP violating phase $\delta $ arises in
these models due to both the mixing of right-handed quarks and the
mixing of left- and right-handed gauge bosons (see Ref.
\cite{Doi}, Eq. (3.1.11)). The parameters $\kappa$, $\eta$, and
$\lambda$ characterize the strength of nonstandard effects. Below,
we give some illustrative examples relating the couplings
$\epsilon^{V-A}_{V-A,i}$, $\epsilon^{V+A}_{V\pm A,i}$ and the
particle masses, couplings and the mixing parameters in the
underlying theoretical models.

\vspace*{3mm} In the R-parity-violating (RPV) SUSY accompanying
the neutrino exchange
mechanism~\cite{Mohapatra86,Vergados87,SUSY1,Babu95,SUSY,Faessler97},
SUSY particles (sleptons, squarks) are present in one of the two
effective 4-fermion vertices. (The other vertex contains the usual
$W_L$ boson.) The nonzero parameters are
\begin{eqnarray}\label{SUSY}
&\ds   \epsilon^{V-A}_{V-A,i}=\frac{1}{2}\eta^{n1}_{(q)RR}U_{ni},\
\
   \epsilon^{S-P}_{S+P,i}=2\eta^{n1}_{(l)LL}U_{ni},\nonumber \\
&\ds   \epsilon^{S+P}_{S+P,i}=-\frac{1}{4}\left(\eta^{n1}_
   {(q)LR}-4\eta^{n1}_{(l)LR}\right)U^*_{ni},\
   \epsilon^{T_R}_{T_R,i}=\frac{1}{8}\eta^{n1}_{(q)LR}U^*_{ni},
\end{eqnarray}
where the index $n$ runs over $e$, $\mu$, $\tau$ (1, 2, 3), and
the RPV Minimal Supersymmetric Model (MSSM) parameters $\eta$s
depend on the couplings of the RPV MSSM superpotential, the masses
of the squarks and the sleptons, the mixings among the squarks and
among the sleptons. Concentrating on the dominant contributions
$\epsilon^{S+P}_{S+P,i}$ and $\epsilon^{T_R}_{T_R,i}$ (as the
others are helicity-suppressed), one can express $\eta^{n1}_{(q)
LR}$ and $\eta^{n1}_{(l) LR}$ as follows~\cite{SUSY}
\begin{eqnarray}\label{SUSY-Etas}
&\ds \eta^{n1}_{(q)LR} =
 \sum_k \frac{\lambda^\prime_{11k} \lambda^\prime_{nk1}}{2\sqrt{2}G_F}
  \sin 2 \theta^d_{(k)} \left(\frac{1}{m^2_{\tilde{d}_1(k)}}
  -  \frac{1}{m^2_{\tilde{d}_2(k)}}   \right),\nonumber \\
&\ds  \eta^{n1}_{(l)LR} =
  \sum_k \frac{\lambda^\prime_{k11} \lambda_{n1k}}{2\sqrt{2}G_F}
  \sin 2 \theta^e_{(k)} \left(\frac{1}{m^2_{\tilde{e}_1(k)}}
  -  \frac{1}{m^2_{\tilde{e}_2(k)}}   \right),
\end{eqnarray}
where $k$ is the generation index, $\theta^d_{(k)}$ and
$\theta^e_{(k)}$ are the squark and slepton mixing angles,
respectively, $ m_{\tilde{f}_1}$ and $ m_{\tilde{f}_2}$ are the
sfermion mass eigenvalues, and $\lambda_{ijk}$ and
$\lambda^\prime_{ijk}$ are the RPV-couplings in the
superpotential.

For the mechanism with LQs in one of the effective vertices
\cite{LQ}, the nonzero coefficients are
\begin{eqnarray}\label{LQ1}
&\ds
\epsilon^{S+P}_{S-P}=-\frac{\sqrt{2}}{4G_F}\frac{\epsilon_V}{M_V^2},\
   \ \epsilon^{S+P}_{S+P}=-\frac{\sqrt{2}}{4G_F}\frac{\epsilon_S}{M_S^2},
   \nonumber \\
&\ds
\epsilon^{V+A}_{V-A}=-\frac{1}{2G_F}\left(\frac{\alpha_S^{(L)}}{M_S^2}
+\frac{\alpha_V^{(L)}}{M_V^2}\right),\ \
\epsilon^{V+A}_{V+A}=-\frac{\sqrt{2}}{4G_F}\left(\frac{\alpha_S^{(R)}}{M_S^2}
+\frac{\alpha_V^{(R)}}{M_V^2}\right),
\end{eqnarray}
where
\begin{equation}\label{eps}
\epsilon^\beta_\alpha=U_{ei}\epsilon^\beta_{\alpha i},
\end{equation}
the parameters $\epsilon_{S(V)}$, $\alpha_{S(V)}^{(L)}$,
$\alpha_{S(V)}^{(R)}$ depend on the couplings of the
renormalizable LQ-quark-lepton interactions consistent with the SM
gauge symmetry, the mixing parameters and the common mass scale
$M_{S(V)}$ of the scalar (vector) LQs~\cite{BRW87}.

The upper bounds on some of the $\epsilon^\beta_\alpha$ parameters
(\ref{eps}) from the Heidelberg--Moscow experiment were derived in
Ref. \cite{0002109} using the $s$-wave approximation for the
electrons, considering nucleon recoil terms and only one nonzero
parameter $\epsilon^\beta_{\alpha i}$ in the Lagrangian at a time,
see Table 1. \vspace*{1mm}
\begin{center}
{\small\noindent Table 1: Upper bounds on some of
$\epsilon_\alpha^\beta$ parameters (\ref{eps}) (CL = 90\%).}\\
\vspace*{1mm}
\begin{tabular}{|c|c|c|c|c|c|}
  \hline
  $\epsilon^{V+A}_{V+A}$ & $\epsilon^{V+A}_{V-A}$ & $\epsilon^{S+P}_{S+P}$ &
  $\epsilon^{S+P}_{S-P}$ & $\epsilon^{T_R}_{T_R}$ & $\epsilon^{T_R}_{T_L}$ \\
  \hline
  $6\times 10^{-7}$ & $4\times 10^{-9}$ & $9\times 10^{-9}$ & $9\times 10^{-9}$ &
  $1\times 10^{-9}$ & $6\times 10^{-10}$ \\
  \hline
\end{tabular}
\end{center}
\vspace*{3mm}

The coefficients $\epsilon_{\alpha i}^{\beta}$ entering the
Lagrangian (\ref{L}) can be expressed as
\begin{equation}
\label{hat}
\epsilon _{\alpha i}^\beta   = \hat \epsilon _\alpha
^\beta \hat U_{ei},
\end{equation}
where $\hat U_{ei}$ are mixing parameters for non-SM neutrinos
(see, e.g., Eq. (\ref{Doi})). As this Lagrangian describes also
ordinary $\beta$-decays (without LN violation), the coefficients
$\hat \epsilon _\alpha ^\beta$ are constrained by the existing
data on precision measurements in allowed nuclear beta decays,
including neutron decay \cite{sev}. For example, from these data
we obtain the conservative bound
\begin{equation}
\label{RR}
\left|{\hat \epsilon _{V + A}^{V + A}}\right| < 7
\times 10^{- 2}.
\end{equation}
From Eqs. (\ref{eps}), (\ref{hat}), (\ref{RR}) and Table 1 we can
assume that the nonstandard mixing is small:
\begin{equation}
\label{nsm}
\left| {U_{ei} \hat U_{ei} } \right|\lesssim 10^{- 5}.
\end{equation}

\subsection{Approximations and electron angular distribution}
We have calculated only the leading order in the Fermi constant
and the leading contribution of the parameters
$\epsilon_\alpha^\beta$ to the decay width using the approximation
of relativistic electrons and non-relativistic nucleons. Following
Ref. \cite{Shchep} we describe the outgoing electrons by plane
waves approximately taking into account the effect of the nuclear
Coulomb field by the Fermi factors $F(\varepsilon_s)$
\cite{Shchep,Doi} with the $s$-th electron energy $\varepsilon_s$.
The non-relativistic structure of the nucleon currents in the
impulse approximation is taken from Ref. \cite{Tomoda}. We neglect
the nucleon recoil terms, because the accurate calculation of the
corrections should also include simultaneously a precise
calculation of the effect of the nuclear Coulomb field, requiring
the technique of the spherical waves for the electrons. Note that
in Ref. \cite{Tomoda} the recoil terms due to the weak magnetism
were calculated in the model with only $V\mp A$ currents (the
recoil terms due to the pseudoscalar form factor were not taken
into account) and it was shown that in the interaction
proportional to $\epsilon_{V-A}^{V+A}$ the recoil effect dominates
over other contributions in the $0^+\!\!\rightarrow\!0^+$
transition. However, our numerical calculations carried out in
section 3 are restricted to the coefficients $U_{ei}$ (i.e., the
SM $+$ light Majorana $\nu$s scenario) and
$\epsilon_{V+A,i}^{V+A}$ (light Majorana $\nu$s in the left-right
symmetric models) entering the Lagrangian (\ref{L}), and in both
cases the nucleon recoil effects are not dominant.

We obtain the differential width in $\cos \theta$, where $\theta$
is the angle  between the electron momenta in the rest frame of
the parent nucleus in the
 $0^+\!(A,Z)\rightarrow\!0^+\!(A,Z+2) e^- e^-$ transitions,
\begin{eqnarray}
\label{dG}
& \ds \frac{d\Gamma}{d\cos\theta}=C|M_{GT}|^2I[(a+b)(1-k\cos\theta)]~,
\nonumber\\
& I[x]=\int d\varepsilon_1\varepsilon_1^2\varepsilon_2^2F
    (\varepsilon_1)F(\varepsilon_2)x~,
\end{eqnarray}
where $\varepsilon_2=\Delta-\varepsilon_1$ and $\Delta$ is the
energy release in the process. The constant
\begin{equation}\label{C}
    C=\frac{G_F^4g_A^4m_e^2}{64\pi^5R_0^2}
\end{equation}
contains the electron mass $m_e$ and the nuclear radius $R_0$, included
in the definition of $C$ so that
the $a$ and $b$ functions and the neutrino potentials are
dimensionless. The Gamow--Teller nuclear matrix element,
\begin{equation}\label{M_GT}
  M_{GT}=\langle0^+_f||\sum\limits_{a\neq
    b}h(r_{ab},\omega){\bm\sigma}_a\cdot{\bm\sigma}_b\tau^a_+\tau^b_+||0^+_i\rangle,
\end{equation}
contains the neutrino potential $h(r,\omega)=R_0\phi_0/r$ with
$\phi_0=e^{i\omega r}$, $r=r_{ab}$ is the distance between the
nucleons $a$ and $b$, and $\omega$ is the average energy of the
 neutrino. The
operator $\tau^a_+=(\tau_1+i\tau_2)^a/2$ converts the $a$-th
neutron into the $a$-th proton; $|0^+_i\rangle$ ($\langle0^+_f|$)
is the initial (final) nuclear state. The angular correlation
coefficient in Eq. (\ref{dG}) is
\begin{equation}\label{k}
k=\frac{a-b}{a+b}~,\quad -1<k\leq 1.
\end{equation}

The expressions for $a$ and $b$ for different choices of
$\epsilon_\alpha^\beta$, considered only one at a time, are shown in Table 2.
In this table
$\varepsilon_{12}=\varepsilon_1-\varepsilon_2$ and $\langle m\rangle$ is
 the effective Majorana mass. The
form factors $g_V(q^2)$, $g_A(q^2)$, $F_S^{(3)}(q^2)$, and
$T_1^{(3)}(q^2)$ describe the following
 nucleon matrix elements \cite{Adler}
\begin{equation}\label{C:scalar}
    \langle P(k^\prime)| \bar ud| N(k) \rangle=
    F^{(3)}_S(q^2)\bar\psi(k^\prime)
    \tau_+ \psi(k),
\end{equation}
\begin{equation}\label{C:vector}
    \langle P(k^\prime)| \bar u 2\gamma^\mu P_{L,R}d| N(k) \rangle=
    \bar\psi(k^\prime)\gamma^\mu\left[g_V(q^2)\mp
    g_A(q^2)\gamma_5\right]
    \tau_+ \psi(k),
\end{equation}
\begin{equation}\label{C:tenzor}
    \langle P(k^\prime)| \bar u2\sigma^{\mu\nu} P_{L,R}d| N(k) \rangle=
    \bar\psi(k^\prime)\left[T_1^{(3)}(q^2)\sigma^{\mu\nu}\mp
    \frac{i}{2}\epsilon^{\mu\nu\rho\sigma}T_1^{(3)}(q^2)\sigma_{\rho\sigma}\right]
    \tau_+\psi(k),
\end{equation}
where
\begin{equation}
    \psi=\left(%
\begin{array}{c}
  P \\
  N \\
\end{array}%
\right)
\end{equation}
is a nucleon isodoublet. We neglect the dipole dependence of the
form factors on the momentum transfer $q=k^\prime-k$ and omit the
zero argument of the form factors. Notations similar to the ones
in Ref.~\cite{Doi} are used in Table 2.

\vspace*{2mm}

\begin{center}
{\small\noindent Table 2: Expressions for $a$ and $b$ in Eqs.
(\ref{dG}) and (\ref{k}) for the stated choice of
$\epsilon_\alpha^\beta$.} \label{Tab1}
\nopagebreak
\begin{tabular}{|c|c|c|}
  \hline
   $\epsilon$ & $a$ & $b$ \\
  \hline
  $\epsilon^{V-A}_{V-A}$ & $|[\langle m\rangle/m_e+2\sum\nolimits_{i}U_{ei}\epsilon^
  {V-A}_{V-A,i}(m_i/m_e)](1-\chi_F)|^2$ & 0 \\
  \hline
  $\epsilon^{V-A}_{V+A}$ & $|(1-\chi_F)\langle
  m\rangle/m_e
  -2\sum\nolimits_{i}U_{ei}\epsilon^{V-A}_{V+A,i}(m_i/m_e)(1+\chi_F)|^2$ & 0 \\
  \hline
  $\epsilon^{S+P}_{S\mp P}$ & $a_0+\frac{1}{9}(\Delta/m_e)^2(F_S^{(3)}/g_V)^2|
   \epsilon^{S+P}_{S\mp P}\chi_F^\prime|^2$ & 0 \\
  \hline
  $\epsilon^{T_R}_{T_L}$ & $a_0
  +\frac{16}{81}(T_1^{(3)}/g_A)^2(
  \Delta/m_e)^2|\epsilon^{T_R}_{T_L}(\chi_{GT}^{\prime}+3\chi_{T}^{\prime})|^2$ & 0 \\
  \hline
  $\epsilon^{T_L}_{T_L}$, $\epsilon^{T_R}_{T_R}$ & $a_0$ & 0 \\
  \hline
  \hline
  $\epsilon$ & $b$ & $a$ \\
  \hline
  $\epsilon^{V+A}_{V-A}$ & $\frac{1}{2}|\epsilon^{V+A}_{V-A}|^2
  [(\varepsilon_{12}/m_e)^2|\chi_{2+}|^2
  +\frac{4}{9}(\Delta/m_e)^2|\chi_P^\prime|^2]$ & $a_0$ \\
  \hline
  $\epsilon^{V+A}_{V+A}$ & $\frac{1}{2}(\varepsilon
  _{12}/m_e)^2|\epsilon^{V+A}_{V+A}\chi_{2-}|^2$ & $a_0$ \\
  \hline
  $\epsilon^{S-P}_{S\mp P}$ & $2(F_S^{(3)}/g_V)^2
  |\sum\nolimits_{i}U_{ei}\epsilon_{S\mp
    P,i}^{S-P}(m_i/m_e)\chi_F|^2$ & $a_0$ \\
  \hline
  $\epsilon^{T_L}_{T_R}$ &
  $32(T_1^{(3)}/g_A)^2|\sum\nolimits_{i}U_
  {ei}\epsilon^{T_L}_{T_R,i}(m_i/m_e)|^2$ & $a_0$ \\
  \hline
\end{tabular}
\end{center}
\vspace*{3mm}
\noindent Thus,
\begin{eqnarray}
&\ds
\chi_{2\pm}=\chi_{GT\omega}\pm\chi_{F\omega}-\frac{1}{9}\chi_{1\mp},\quad
\chi_{1\pm}=\chi^\prime_{GT}-6\chi_T^\prime\pm3\chi_F^\prime,\\
&\ds \chi_F=\left(\frac{g_V}{g_A}\right)^2\frac{M_F}{M_{GT}},\quad
\chi_P=\left(\frac{g_V}{g_A}\right)\frac{M_P}{M_{GT}},\quad
\chi_X=\frac{M_X}{M_{GT}}, \ X=T,\,GT,
\end{eqnarray}
with Fermi $M_F$, pseudoscalar $M_P$, and tensor $M_T$ nuclear
matrix elements:
\begin{eqnarray}
&\ds   M_{F}=\langle0^+_f||\sum\limits_{a\neq b}h(r_{ab},\omega)
   \tau^a_+\tau^b_+||0^+_i\rangle, \\
&\ds   M_{P}=\langle0^+_f||\sum\limits_{a\neq b}h(r_{ab},\omega)
   \left\{(\bm\sigma_a-\bm\sigma_b)\cdot\left[{\bm n}\times
   {\bf n_+}\right]\right\}\tau^a_+\tau^b_+||0^+_i\rangle, \\
&\ds   M_{T}=\langle0^+_f||\sum\limits_{a\neq b}h(r_{ab},\omega)
   \left[{\bm\sigma}_a\cdot{\bm n}{\bm \sigma}_b\cdot{\bf n}-\frac{1}{3}{\bm\sigma}
   _a\cdot{\bm\sigma}_b\right]\tau^a_+\tau^b_+||0^+_i\rangle,
\end{eqnarray}
with
\begin{equation}
{\bf r}=r{\bf n}, \qquad {\bf R}=R{\bf n}_+,
\end{equation}
where ${\bf r}={\bf r}_{ab}$ (${\bf R}={\bf R}_{ab}$) is the
difference (half sum) of radius-vectors of the nucleons $a$ and
$b$; ${\bf n}$ and ${\bf n}_+$ are unit vectors. The prime and the
index $\omega$ imply that the matrix element in the numerator
instead of $h$ contains the neutrino potential $h^\prime=h+\omega
R_0h_1$ or $h_\omega=h-\omega R_0h_1$, respectively, with
${h_1=-d\phi_0/d(\omega r)}$. The quantity $a$ for all zero
$\epsilon_\alpha^\beta$ is called $a_0$ in Table 1, and is defined
as:
\begin{equation}\label{a0}
    a_0\equiv|(1-\chi_F)\langle m\rangle/m_e|^2.
\end{equation}
For the $V\mp A$ part of the Lagrangian (\ref{L}), our result
agrees with Ref. \cite{Doi} for the relativistic electrons
($m_e/\Delta\to0$) that weakly interact with the nucleus ($\alpha
Z\to 0$), if the recoil and P-wave effects are not taken into
account.

If the effects of all the interactions beyond the SM extended by
the $\nu_M$s, which we call the ``nonstandard" effects, are zero
(i.e., all $\epsilon^\beta_\alpha=0$), then $k = 1$  and the
distribution (\ref{dG}) is proportional to $1-\cos\theta$. The
angular coefficient deviates from 1 only for the cases $b\neq 0$
irrespective of the value of $a$. Therefore, the presence of the
``nonstandard" {\it first} set of parameters in Table~2,
$\epsilon^{V-A}_{V\mp A}$, $\epsilon^{S+P}_{S\mp P}$,
$\epsilon^{T_R}_{T_{L}}$, $\epsilon^{T_L}_{T_L}$ and
$\epsilon^{T_R}_{T_R}$ does not change the form of the angular
distribution, but the presence of the {\it second} set (see lower
part of Table~2), $\epsilon^{V+A}_{V\mp A}$, $\epsilon^{S-P}_{S\mp
P}$, $\epsilon^{T_L}_{T_R}$, $\epsilon^{T_L}_{T_L}$ and
$\epsilon^{T_R}_{T_R}$ does change this distribution. Thus,
experimentally establishing $k \neq 1$ would signal the presence
of beyond-the-SM contribution in the $0\nu2\beta$ decay. The
converse is not true; namely establishing $k = 1$ experimentally
will not single out the SM + $\nu_M$s as the only mechanism of the
$0\nu2\beta$ decay and one would require additional
input/constraints on the parameters of the
underlying theory with their coefficients
$\epsilon_{\alpha}^{\beta}$ listed in the upper part of Table 2.
The coefficient $k$ and the set $\{\epsilon\}$ of nonzero
$\epsilon_\alpha^\beta$s that change the
 $1-\cos\theta$ form of the distribution for the SM plus $\nu_M$s
are given in Table 3 (the lower two entries). They correspond to
the following extensions of the SM:  $\nu_M$s plus RPV SUSY
\cite{SUSY}, $\nu_M$s plus right-handed currents (RC) (connected
with right-handed $W$ bosons \cite{Doi} or LQs \cite{LQ}).
\begin{center}
{\small \noindent Table 3: The angular correlation coefficient $k$
for various SM extensions.} \vspace*{1mm}

\begin{tabular}{|c|c|c|}
  \hline
  SM extension & $\{\epsilon\}$ & $k$ \\
  \hline
  $\nu_M$ & --- & 1 \\
  \hline
  $\nu_M$+RPV SUSY & $\epsilon_{S+P}^{S-P}$ & $1-k \ll 1$ \\
  \hline
  $\nu_M+$RC & $\epsilon_{V\mp A}^{V+A}$ & $-1<k\leq 1$ \\
  \hline
\end{tabular}
\end{center}

Among the models that we have listed in Table 3, the coefficient
$\epsilon_{S+P}^{S-P}$ for the $\nu_M +$ RPV SUSY case is helicity
suppressed (and $\epsilon_{S-P}^{S-P} = 0$), and hence the angular
coefficient $k\simeq 1$ for this model. (It is a mathematical
challenge to come up with a model which lifts this chiral
suppression).
 Among the realistic models we have discussed, only
the model called $ \nu_M +$ RC can essentially change the angular
coefficient $k$ from being 1. Left-right symmetric models belong
to this class and we have studied these models in detail in
section 3, where the correlation among the parameters $K$ (see Eq.
(\ref{tild k}) below), $m_ {W_R}$ and $\left|\langle m
\rangle\right|$ (or $T_{1/2}$) is worked out.

For a quantitative understanding of the influence of the decay
mechanism on the angular distribution, it is necessary to take
into account the effect of the nuclear Coulomb field on the
electrons in the terms with $\epsilon_{S\mp P}^{S\mp P}$ ~and
$\epsilon_{T_{L,R}}^{T_{L,R}}$ and  the nucleon recoil terms. Note
that the calculation of the above-mentioned corrections for the
long range mechanism of the $0\nu2\beta$ decay will change
inessentially the values of $a$ and $b$ for
$\epsilon_{V-A}^{V-A}$, $\epsilon_{V+A}^{V-A}$ or
$\epsilon_{V+A}^{V+A}$, varied one  at a time, and hence the
results of the next section.

\section{Electron angular correlation in left-right symmetric
models}

The experimental bounds on the $\epsilon_\alpha^\beta$ are
connected with the masses of new particles, their mixing angles,
and other parameters specific to particular extensions of the SM
\cite{Doi,Shchep,SUSY,SUSY1,BL,LQ}. To illustrate the kind of
correlations that the measurements of  $|\langle m\rangle|$ and
the angular correlation coefficient $k$ in the $0\nu2\beta$ decay
would imply, we work out the case of the left-right symmetric
models \cite{LR}. In the model $SU(2)_L\times SU(2)_R\times U(1)$
the parameter $\lambda$ (see Eq. (\ref{Doi})) is expressed through
the masses $m_{W_{L}}$ and $m_{W_{R}}$ of the left- and
right-handed $W$ bosons\cite{Doi}:
\begin{equation}\label{r1}
\lambda  = \left( {m_{W_L } /m_{W_R } } \right)^2 ,
\end{equation}
under the condition
\begin{equation}\label{mL<<mR}
 m_{W_L}\ll m_{W_R}.
\end{equation}
Eqs. (\ref{Doi}) and (\ref{eps}) yield the relation
\begin{equation}\label{r2}
\epsilon^{V+A}_{V+A}=\lambda\frac{g^\prime_V}{g_V}U_{ei}V_{ei}.
\end{equation}
\noindent To reduce the number of free parameters, we assume the
equality of the form factors of the left- and right-handed
hadronic currents:
\begin{equation}\label{gV}
g_V=g^\prime_V.
\end{equation}
The small masses of the observable $\nu$s are likely described by
the seesaw formula that in the simplest case gives
\begin{equation}
    m_i\sim m_D^2/M_R,~~~ M_R\gg m_D,
\end{equation}
with the Dirac mass scale $m_D$ (for the charged leptons and the
light quarks $m_D\gtrsim 1$ MeV) and the mass scale $M_R$ of right
$\nu_M$s (in the majority of theories $M_R\gtrsim 1$ TeV). In the
left-right symmetric models these scales arise usually from the
two scales of the vacuum expectation values of Higgs multiplets
\cite{LR}. In the seesaw mechanism, the values of the mixing
parameters $V_{ei}$ (for $i$ numbering light mass states) have the
same order of magnitude as $m_D/M_R$. In our discussion we use two
rather conservative values (compare with Eq. (\ref{nsm}))
\begin{equation}\label{e}
\epsilon=10^{-6},~ 5\times10^{-7}
\end{equation}
for the mixing parameter
\begin{equation}\label{epsilon}
\epsilon=|U_{ei}V_{ei}|.
\end{equation}
We recall that here the summation index $i$ runs only over the light
neutrino mass eigenstates (the summation over the total mass
spectrum including also heavy states gives strictly zero due to
the orthogonality condition \cite{Doi}).

From Eqs. (\ref{r1}), (\ref{r2}), (\ref{gV}) and (\ref{epsilon})
we have
\begin{equation}
\label{WR} m_{W_R} =
m_{W_L}\left(\epsilon/\left|\epsilon^{V+A}_{V+A}\right|\right)^{1/2}.
\end{equation}
Using Eq. (\ref{mL<<mR}) we note the approximate equality of
$m_{W_{L}}$ and the mass of the observed charged gauge boson $W_1$
($m_{W_1}$=80.4~GeV \cite{PDG}).

We now turn to work out the relations among the angular
correlation coefficient $k$, the right-handed $W$-boson mass
$m_{W_R}$ and the neutrino effective mass $|\langle m\rangle|$. To
this end, we note that the differential width for the only nonzero
nonstandard parameter $\epsilon_{V+A}^{V+A}$ can be obtained by
comparing Eqs.~(\ref{dG}), (\ref{k}), (\ref{a0}) and the
corresponding expression for $b$ from Table 2 with the more
precise result of Ref. \cite{Doi}, yielding:
\begin{equation}\label{dG1}
\frac{d\Gamma}{d\cos\theta}(\epsilon_{V+A}^{V+A})=|M_{GT}|^2\frac{\ln2}{2}
(A+ B)(1- K \cos\theta),
\end{equation}
which is correct for the spherical waves of electrons distorted by
the Coulomb field of the nucleus in the limit of small $m_e/\Delta$.
Here,
\begin{equation}\label{tild k}
K=\frac{A - B}{A + B}~;\quad  A=a_0G_{01},\quad
B=\left|\epsilon^{V+A}_{V+A}\chi_{2-}\right|^2G_{02},
\end{equation}
with the usual phase space factors $G_{0i}$ ($i=1,2$) defined in
Ref.~\cite{Doi}. Note that the angular coefficient $K$ entering in
 Eq.~(\ref{dG1})
differs from the coefficient $k$, entering Eqs.~(\ref{dG}) and
(\ref{k}), as $k$ is a function of the electron energy
$\varepsilon_1$ and $K$ is obtained by integrating over the energy
spectrum in Eq.~(\ref{dG}).
 The $a_0$ and $\chi_{2-}$ in Eq. (\ref{tild
k}) contain instead of $\phi_0$ the neutrino potential $\phi$ from
Ref. \cite{Doi}. In the numerical calculation we have used
$\chi_F=0.274$, $\chi_{2-}=0.551$, $M_{GT}=1.846$, obtained in a
QRPA model with p-n pairing for $^{76}\mbox{Ge}$, with
$G_{01}=7.928\times 10^{-15}$ and $G_{02}=12.96\times 10^{-15}$
(yr$^{-1}$) \cite{Pantis}. For a recent discussion on
uncertainties in $0\nu2\beta$ decay nuclear matrix elements see
Ref. \cite{Tuebingen}.

Using Eqs. (\ref{tild k}), (\ref{a0}), and (\ref{WR}), we have
\begin{equation}
K = \frac{{y - 1}}{{y + 1}},\;y = \frac{{G_{01} }}{{G_{02}
}}\left( {\frac{{1 - \chi _F }}{{\chi _{2 - } \epsilon
}}\frac{{\left| {\left\langle m \right\rangle } \right|}}{{m_e }}}
\right)^2 \left( {\frac{{m_{W_R } }}{{m_{W_L } }}} \right)^4.
\end{equation}
The correlation among  $K$, $m_{W_R}$ and $|\langle m\rangle|$ is
shown in Fig.~1 for $\epsilon=10^{-6}$ and in Fig.~2 for
$\epsilon=5\times 10^{-7}$. We consider the values of $|\langle
m\rangle|$, starting from the current upper bound from the
Heidelberg--Moscow experiment, taken as $|\langle m\rangle|\leq
0.3$~eV, up to $|\langle m\rangle|= 0.001$~eV, covering most
scenarios of neutrino mass hierarchies and mixing angles (see
Ref.~\cite{Bil} for a recent discussion and update). Concerning the
existing bounds on $m_{W_R}$, we
note that from Eqs. (\ref{Doi}), (\ref{hat}) and  (\ref{gV}) one
obtains ${\hat \epsilon _{V + A}^{V + A}} = \lambda$. With this, Eq.
(\ref{r1}) and the constraint (\ref{RR}) derived from \cite{sev}
yield  $m_{W_R} > 300~\mbox{GeV}$. This bound is weaker than
the one $m_{W_R} > 715~\mbox{GeV}$, obtained  from the electroweak
fits \cite{PDG}. There is still a more stringent bound
$m_{W_R} > 1.2~\mbox{TeV}$, obtained in
Ref.~\cite{2004} for the decay mediated by heavy Majorana
neutrinos using arguments based on the vacuum stability
\cite{Mohapatra86}, but it requires additional theory input.
We assume $m_{W_R }\ge 1~\mbox{TeV}$ in all our
figures.

Using (\ref{WR}), (\ref{dG1}), (\ref{tild k}) and the relation
$T_{1/2}=\ln2/\Gamma$ we get the correlation among $m_{W_R}$ and
the measurable $0\nu2\beta$ decay parameters, namely the half-life
$T_{1/2}$ and the angular coefficient $K$:
\begin{equation}
K=1-2G_{02}(|M_{GT}|\chi_{2-}\epsilon)^{2}(m_{W_L}/m_{W_R})^{4}T_{1/2}.
\end{equation}
The correlation among $K$, $m_{W_R}$ and $T_{1/2}$ is shown in
Fig.~3 for $\epsilon=10^{-6}$ and in Fig.~4 for
$\epsilon=5\times10^{-7}$.

Figs.~ 1 -- 4 are the principal numerical results of this paper.
They show that depending on the values of $|\langle m\rangle|$ (or
$T_{1/2}$) and $m_{W_R}$, all values of the angular coefficient
$K$ are allowed. For example, Fig.~1 shows that it is possible to
describe the angular distribution close to $1+\cos\theta$ ($K\gs
-1$) by the (long range) mechanism with the right-handed boson
$W_R$ with the mass about 1~TeV for $|\langle m\rangle|\sim
1$~meV.

For illustration, in Fig. 5 we plot the differential width
(\ref{dG1}) vs. $\cos\theta$ for a set of values of $|\langle
m\rangle|$ and $m_{W_R}$, assuming  $\epsilon=10^{-6}$. It is seen
that the sensitivity of the electron angular distribution to the
right-handed $W$-boson mass $m_{W_R}$ increases with decreasing
values of the effective Majorana neutrino mass $|\langle
m\rangle|$, as can be seen from Fig. 5 (right), where this
distribution is shown for $|\langle m\rangle|=1$~meV, 3~meV and
5~meV.

\section*{Acknowledgments}
We thank Alexander Barabash and Alexei Smirnov for helpful
discussion.

\newpage

\begin{figure}
\vspace{0.6cm}
\hspace{-1.5cm}
\begin{minipage}{0.49\textwidth}
\includegraphics[scale=0.7]{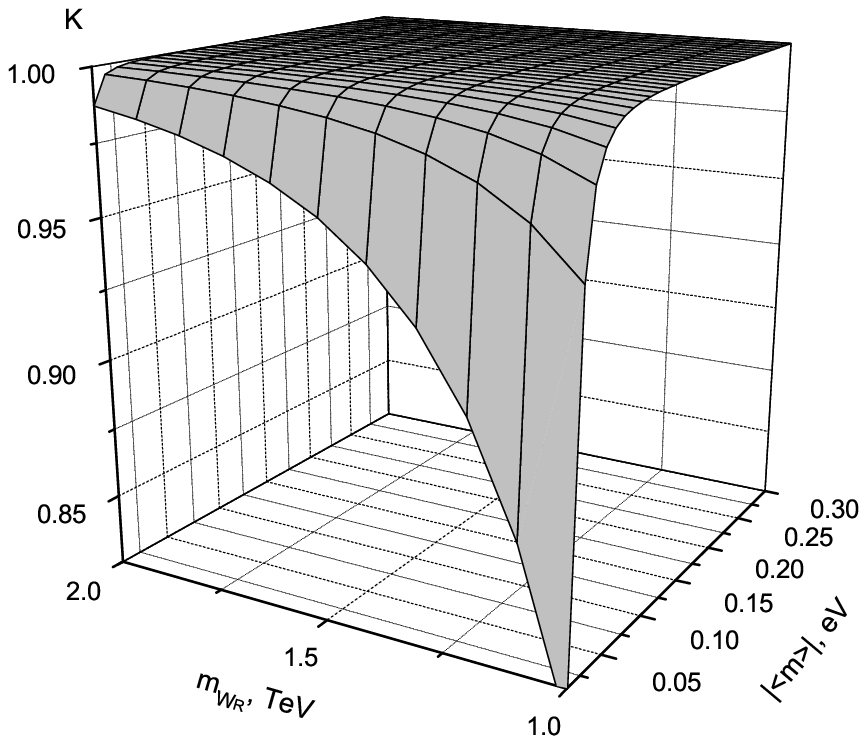}
\end{minipage}
\begin{minipage}{0.49\textwidth}
\includegraphics[scale=0.7]{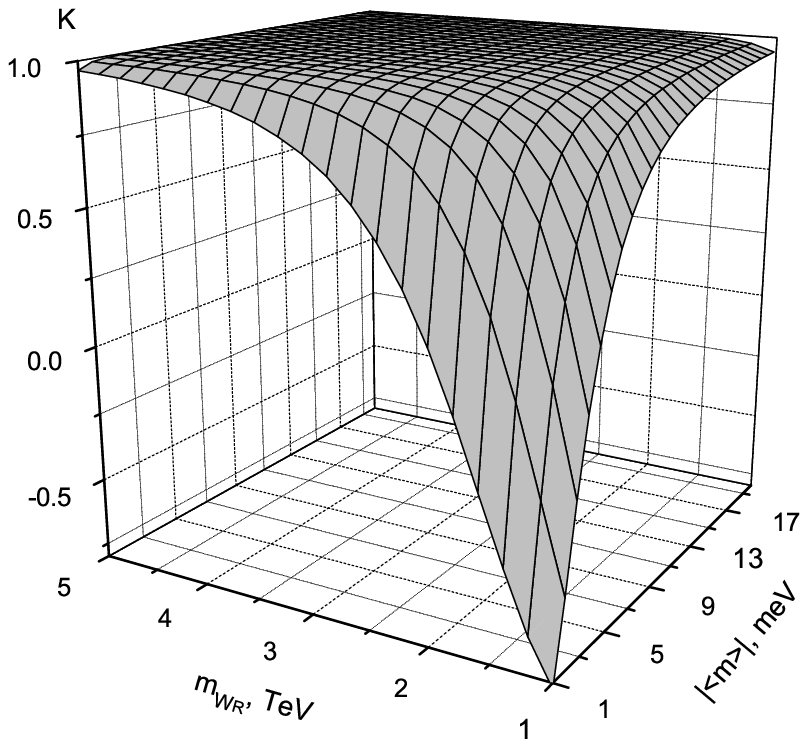}
\end{minipage}
\caption{{\it Left}:~Dependence of the angular correlation
coefficient $K$ on the right-handed $W$-boson mass $m_{W_R}$ and
the value of the neutrino effective mass $|\langle m\rangle|$ for
the $0\nu2\beta$ decay of $^{76}\mbox{Ge}$. The mixing parameter
$\epsilon=10^{-6}$. {\it Right}:~The same as the left figure~but
for smaller values of $|\langle m\rangle|$.} \label{Fig1}
\end{figure}

\begin{figure}
\vspace{0.6cm}
\hspace{-1.5cm}
\begin{minipage}{0.49\textwidth}
\includegraphics[scale=0.7]{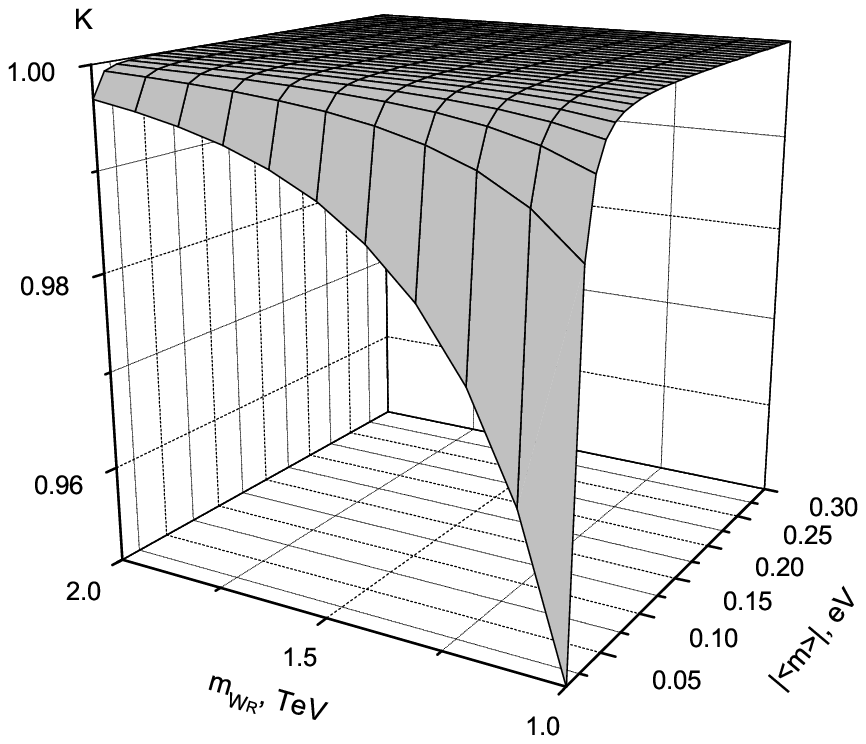}
\end{minipage}
\begin{minipage}{0.49\textwidth}
\includegraphics[scale=0.7]{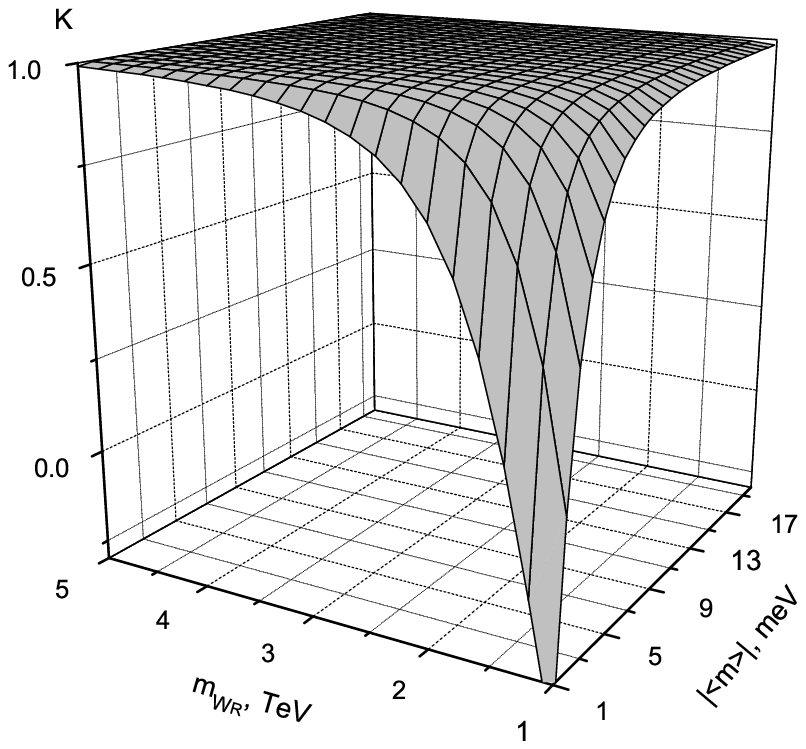}
\end{minipage}
\caption{The same as Fig. 1 ~but for the smaller mixing parameter
$\epsilon = 5\times10^{-7}$.}
\label{Fig2}
\end{figure}

\begin{figure}
\vspace{0.6cm}
\hspace{-1.5cm}
\begin{minipage}{0.49\textwidth}
\includegraphics[scale=0.7]{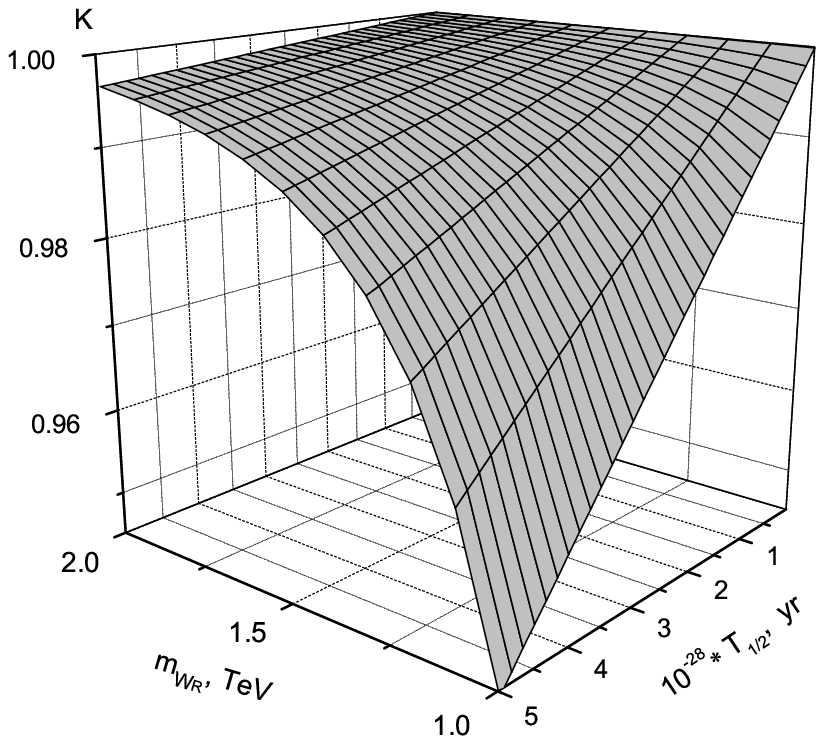}
\end{minipage}
\begin{minipage}{0.49\textwidth}
\includegraphics[scale=0.7]{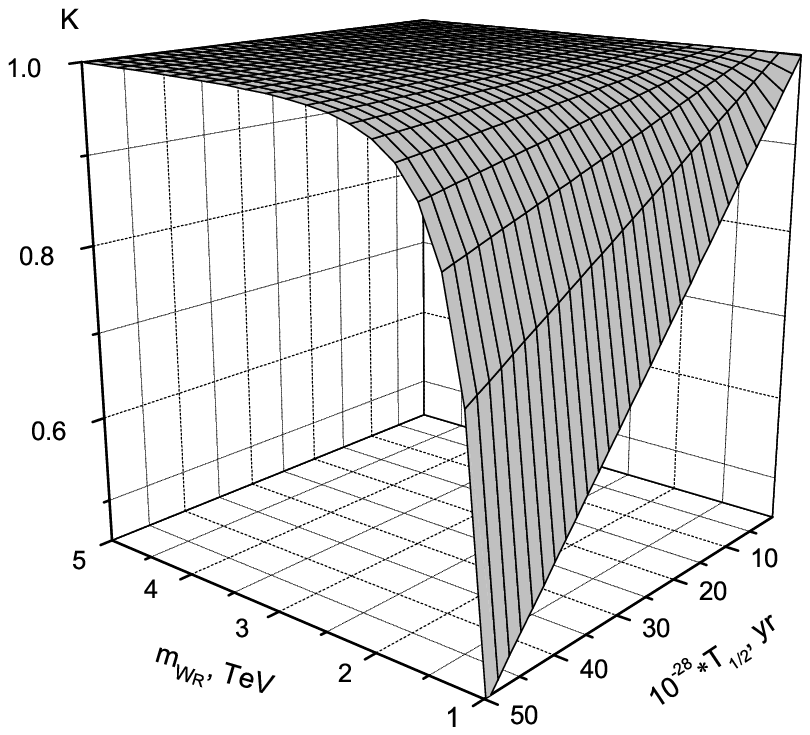}
\end{minipage}
\caption{{\it Left}:~Dependence of the angular correlation
coefficient $K$ on the right-handed $W$-boson mass $m_{W_R}$ and
the half-life $T_{1/2}$ for the $0\nu2\beta$ decay of
$^{76}\mbox{Ge}$. The mixing parameter $\epsilon=10^{-6}$. {\it
Right}:~The same as the left figure~but for larger values of
$T_{1/2}$.} \label{Fig3}
\end{figure}
\begin{figure}
\vspace{0.6cm}
\hspace{-1.5cm}
\begin{minipage}{0.49\textwidth}
\includegraphics[scale=0.7]{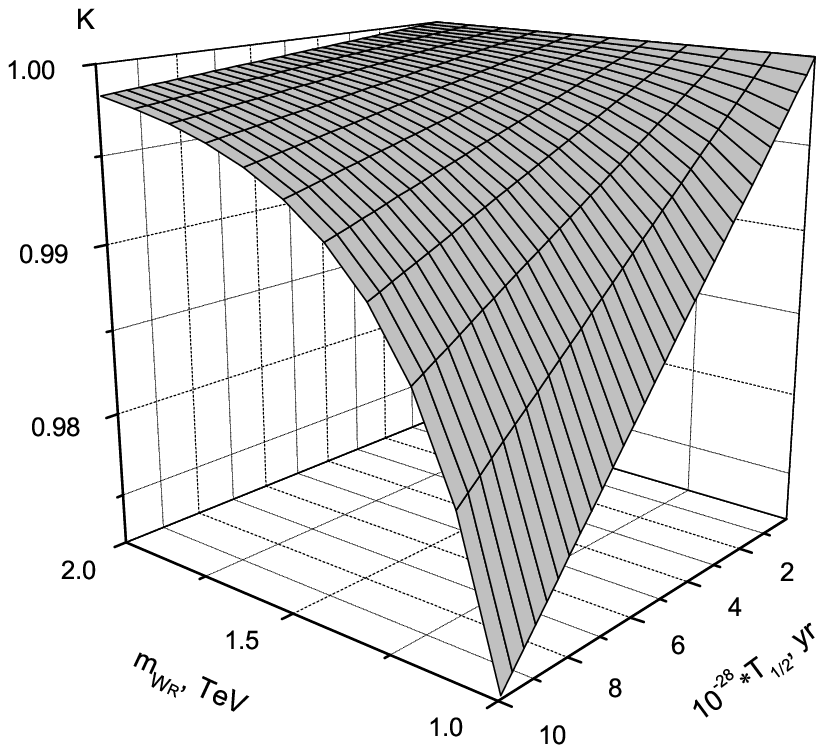}
\end{minipage}
\begin{minipage}{0.49\textwidth}
\includegraphics[scale=0.7]{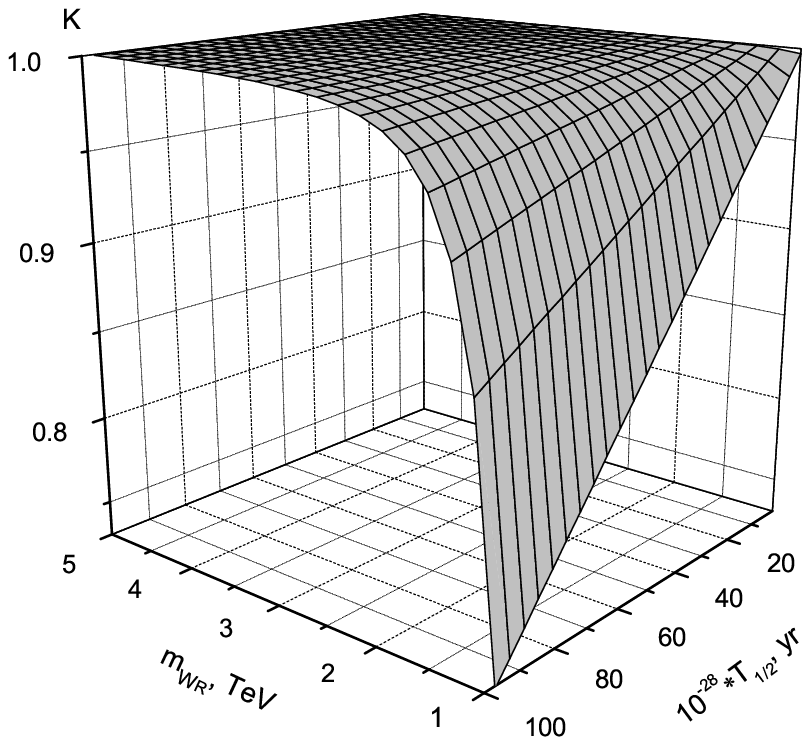}
\end{minipage}
\caption{The same as Fig. 3~but for the smaller mixing parameter
$\epsilon = 5\times 10^{-7}$.} \label{Fig4}
\end{figure}
\begin{figure}
\vspace{-5.0cm}
\hspace{-0.8cm}
\begin{minipage}{0.49\textwidth}
\includegraphics[scale=0.7]{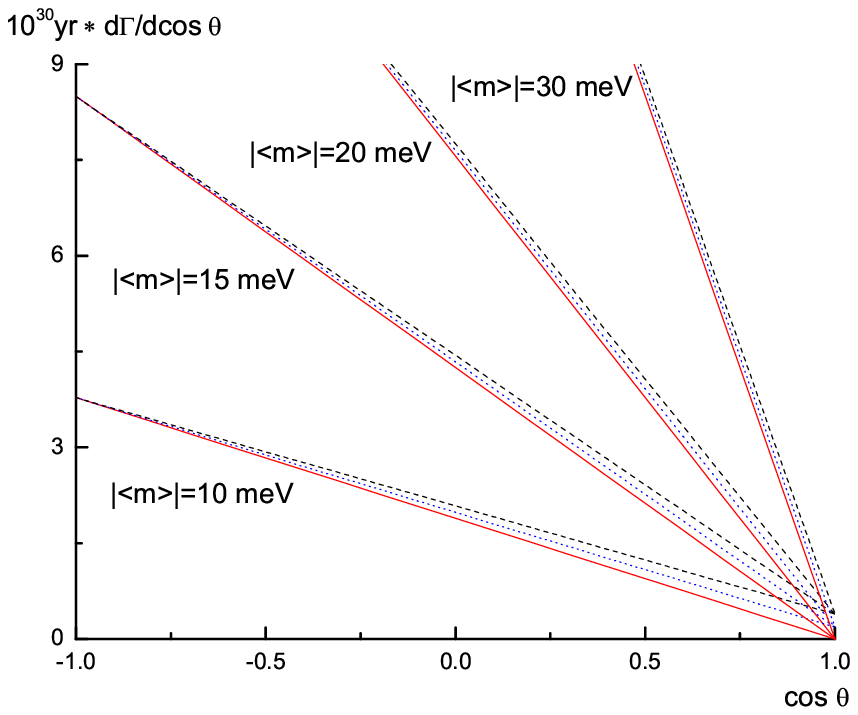}
\end{minipage}
\begin{minipage}{0.49\textwidth}
\includegraphics[scale=0.7]{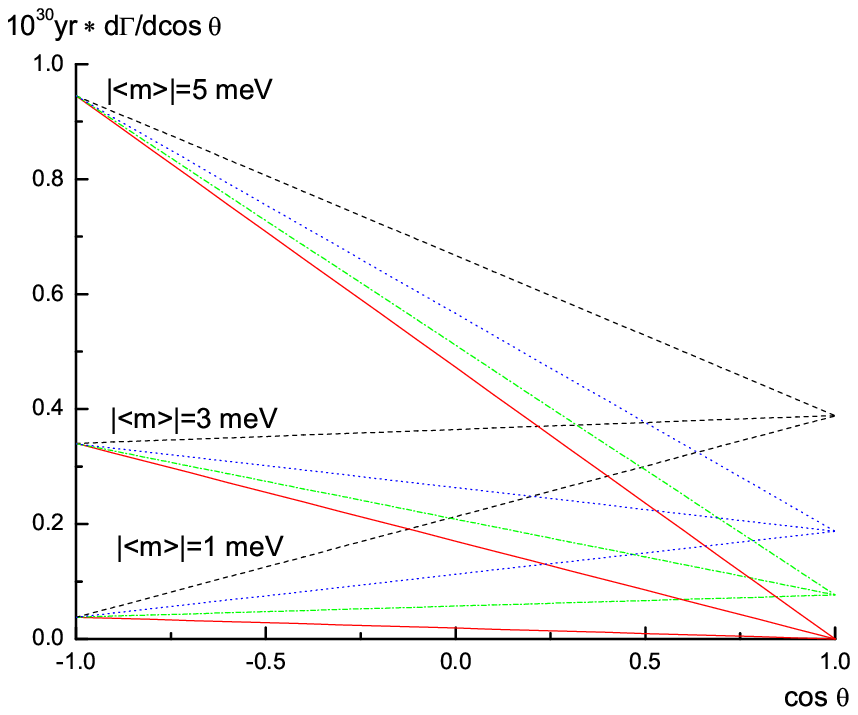}
\end{minipage}
\caption{{\it Left}:~Dependence of the differential width for the
$0\nu2\beta$ decay of $^{76}\mbox{Ge}$ on $\cos\theta$ for a fixed
value of $\epsilon=10^{-6}$ and $|\langle m\rangle|=10, 15, 20, 30
~\mbox{meV}$. The dashed, dotted and straight lines correspond to
$m_{W_R}=1, 1.2, \infty~\mbox{TeV}$, respectively (the latter is
the conventional case of the light Majorana neutrino exchange
mechanism). {\it Right}:~The same as the left figure but for
smaller values of $|\langle m\rangle|= 1, 3, 5~\mbox{meV}$. In
addition, the dash-dotted lines correspond to $m_{W_R}=
1.5~\mbox{TeV}$.} \label{Fig5}
\end{figure}

\end{document}